\documentclass{article}

\usepackage[preprint]{neurips_2025}

\usepackage[utf8]{inputenc}
\usepackage[T1]{fontenc}
\usepackage{hyperref}
\usepackage{url}
\usepackage{booktabs}
\usepackage{amsfonts}
\usepackage{nicefrac}
\usepackage{microtype}
\usepackage{xcolor}
\usepackage{amsmath}
\usepackage{graphicx}

\title{Large Language Models Can Be a Viable Substitute for Expert Political Surveys When a Shock Disrupts Traditional Measurement Approaches}

\author{%
  Patrick~Y.~Wu \\
  Department of Computer Science\\
  American University\\
  Washington, DC 20016 \\
  \texttt{patrickwu@american.edu} 
}

\begin{document}

\maketitle

\begin{abstract}
After a disruptive event or shock, such as the Department of Government Efficiency (DOGE) federal layoffs of 2025, expert judgments are colored by knowledge of the outcome. This can make it difficult or impossible to reconstruct the pre-event perceptions needed to study the factors associated with the event. This position paper argues that large language models (LLMs), trained on vast amounts of digital media data, can be a viable substitute for expert political surveys when a shock disrupts traditional measurement. We analyze the DOGE layoffs as a specific case study for this position. We use pairwise comparison prompts with LLMs and derive ideology scores for federal executive agencies. These scores replicate pre-layoff expert measures and predict which agencies were targeted by DOGE. We also use this same approach and find that the perceptions of certain federal agencies as knowledge institutions predict which agencies were targeted by DOGE, even when controlling for ideology. This case study demonstrates that using LLMs allows us to rapidly and easily test the associated factors hypothesized behind the shock. More broadly, our case study of this recent event exemplifies how LLMs offer insights into the correlational factors of the shock when traditional measurement techniques fail. We conclude by proposing a two-part criterion for when researchers can turn to LLMs as a substitute for expert political surveys.
\end{abstract}

\section{Introduction}

\textbf{This position paper argues that large language models (LLMs) can be a viable substitute for expert political surveys when a shock disrupts traditional measurement approaches.} When shocks, defined as sudden or unexpected events, occur, there is usually a high demand for understanding the magnitude of the shock and what factors potentially contributed or predicted the occurrence of the shock. Even if an expected event occurs, the magnitude of the event may be much broader and disruptive than expected. 

For example, the Department of Government Efficiency (DOGE) layoffs of 2025 are considered a shock \citep{cbsnews_federal_workers_2025}. The DOGE layoffs refer to the mass terminations of American federal government employees across federal executive agencies coordinated by DOGE, an initiative established by Donald Trump's administration in January 2025. Although such job cuts were expected after Donald Trump's presidential electoral victory, the speed and scale of the job cuts are unprecedented (and, as of this paper's writing, are still ongoing and subject to a raft of lawsuits).

Naturally, many scholars have analyzed potential predictors of DOGE layoffs to better understand the factors associated with which agencies were (or will be) targeted. Principally, the firings are considered politically motivated. In other words, the perceived ideology of federal agencies---the perceptions of the policy leanings of the agencies---are largely considered a major predictor of which agencies experience layoffs. \citet{Bonica2025}, using survey data from 2014, empirically demonstrated that the perceived ideology of a federal agency significantly predicted whether it experienced DOGE layoffs, even when controlling for agency staffing and annual budgeting. 

Given that this survey is more than a decade old, current data would help determine if this pattern remains empirically valid. Additionally, examining factors beyond ideology that may predict DOGE layoffs would provide deeper insights into the ongoing firings. For example, many influential far-right figures, some directly cited by officials such as JD Vance, have advocated for the dismantling the ``Cathedral,'' a collective term referring to universities, the media, and civil service \citep{woods2019cultural}.

Measures such as perceived ideology or perception as knowledge institutions are usually estimated using surveys of political experts. For example, \citet{richardson_clinton_lewis_2018} surveyed bureaucrats to estimate the perceived ideology of federal agencies. However, obtaining valid estimates during or after an event such as the DOGE layoffs using such an approach may be impossible, as it would require pre-event perceptions of these federal agencies. Estimating a measure using surveys now, after the start of the DOGE layoffs, would bias downstream inferences that seek to explain which agencies were targeted.

We contend that LLMs are a potential, though imperfect, solution to this problem. We focus on the DOGE layoffs shock as a case study. We prompt pairwise comparisons with several LLMs and derive ideology scores for federal agencies. The outcomes of these pairwise comparisons are then used with the Bradley-Terry model to estimate an ideology score for each agency. We demonstrate that this approach works with three open weight LLMs of varying sizes: each measure replicates pre-layoff expert measures and confirms \citet{Bonica2025}'s finding that ideology predicts which agencies were targeted by DOGE, even when controlling for agency staffing and budget size.\footnote{To be clear, we make no claims about causation; our analysis identifies factors that are associated with which agencies experienced layoffs.}

Based on influential far-right thought from individuals such as Curtis Yarvin, we also use the same approach above to estimate the perception of each agency as a \textit{knowledge institution}, or institutions perceived to produce, distribute, or legitimize knowledge. This measure has not been estimated before for federal agencies; a pre-shock measure is impossible to estimate ex post. We show that this knowledge institution measure predicts which agencies are hit with DOGE layoffs, even when holding constant ideology measures, staff size, and budget size. 

This new measure suggests that LLMs are a viable substitute for expert political surveys when a shock disrupts traditional measurement approaches. LLMs are also much faster than expert surveys. This allows researchers to rapidly test and refine analyses without the delays and costs associated with traditional survey methods. This speed is particularly advantageous when studying rapidly evolving situations, such as the implications of the DOGE layoffs, where timely analyses can inform our understanding of potential institutional vulnerabilities and authoritarian pressures before they fully manifest into threats to constitutional governance and democracy.

This paper is organized as follows. We first review how latent attributes of political actors are typically measured and how they can be measured using LLMs. We then conduct our analyses of the DOGE layoffs as outlined above. We use this case study to discuss the potential advantages and pitfalls of our argued position. Lastly, we propose a two-part theory-gap criterion for when researchers could use an LLM in place of expert political surveys: there must be a substantive theory---real-world evidence that the construct should matter for the outcome of interest---and that there must be a measurement gap---expert survey data cannot be collected now, but it must have been theoretically possible to collect this data before the shock.   

\section{Measuring Latent Attributes of Political Actors}
\subsection{Latent Attributes, Measurement, and Scaling in Political Science}
In political science, latent positions, attributes, or constructs (we use these three terms interchangeably) refer to underlying and unobservable characteristics or stances of political actors, such as legislators, voters, parties, or agencies. These attributes or stances are typically expressed as a scalar value (or vector for multidimensional latent spaces). The predominant focus of latent attribute measurement in political science has been on ideological positioning along the liberal-conservative spectrum. The most well-known example of a liberal-conservative ideology measure is NOMINATE, a multidimensional scaling method that spatially maps congressional voting patterns \citep{keith_poole_nominate,carroll_lewis_lo_poole_rosenthal}. The resulting two-dimensional map represents legislators as points within this ideological space. Across all congresses, the first dimension is interpreted as liberal-conservative ideology: legislators with lower scalar values on this dimension are more liberal, while legislators with higher scalar values are more conservative. The interpretation of the second dimension differs by time period, representing positions on different contemporaneous issues.

As hinted above, latent attributes are not limited to liberal-conservative ideology. Examples of measures that go beyond liberal-conservative ideology include measures that position political actors along policy-specific dimensions \citep{wu2023largelanguagemodelsused}, identify levels of political sophistication \citep{luskin1987sophistication}, and quantify persuasive capabilities \citep{LOEWEN2012212}.

Understanding latent positions is useful because it reduces the dimensionality of political actors' complex actions and stances to a low-dimensional (often unidimensional) scale. We can use this to better understand patterns in political behavior, such as identifying the fundamental dimensions that structure political decisions and actions \citep[see, e.g.,][]{poole1997ideology}. These measures can also be combined with other data to better understand politics and the behaviors of political actors. For example, latent ideological measures have been used to understand how lawmakers represent constituents \citep[see, e.g.,][]{ansolabehere_snyder_stewart_2001,bartels_2009} or how position taking occurs outside of roll call voting \citep[see, e.g.,][]{boudreau_etal_2019,russell_2021,Macdonald2025moderate}.

While there is broad agreement that political actors have positions in these latent low-dimensional spaces, we cannot directly observe these positions. We must use various scaling techniques to estimate these positions using (often imperfect) observed variables. Broadly, observed variables can be categorized into behavior-based variables or perception-based variables. Behavior-based measures use political actors' actions and interpret these actions as revealed preferences. Examples of these actions include decisions on roll call votes \citep{keith_poole_nominate,carroll_lewis_lo_poole_rosenthal}, what words were chosen in a floor speech \citep{gentzkow2019speech}, and which links are shared on social media \citep{Eady_Bonneau_Tucker_Nagler_2025}. Perception-based measures use the perceptions of political actors as the observed variable to estimate latent positions. For example, liberal-conservative measures based on campaign contributions assume ideological homophily between donors and candidates; in other words, it is based on the perceptions of donors \citep{bonica2013}. Similarly, under the same homophily assumption, someone choosing to follow the account of a particular political candidate or organization on social media indicates some kind of political perception of that account \citep{barbera_2015}. 

\subsection{Measuring Latent Attributes of Federal Agencies}

Latent attributes or constructs are not only properties of individuals (e.g., members of Congress). Political parties, interest groups, nonprofit organizations, and federal agencies are all collectives that can also have latent attributes estimated. Almost all studies of latent attributes of federal agencies have focused on two attributes: liberal-conservative ideology and agency independence \citep{Carpenter_2001}. Here, we focus on the former attribute.

One of the difficulties of studying the ideology of federal agencies is that it is unclear where ``ideology'' comes from. Political appointees of federal agencies change with administrations, but career bureaucrats often work through several administrations. Many ways of measuring the liberal-conservative ideology of agencies have been proposed. Some have coded agencies based on whether they were created by a Democratic or Republican president \citep{gilmour2006politicalappointees}. But this is not particularly accurate, as issues of interest within parties shift over time \citep{clinton2012separatedpowers}. For example, the Environmental Protection Agency (EPA) was proposed and established by Richard Nixon, a Republican president. 

Several approaches using observed behaviors have also been proposed. Most of these approaches scale the votes of commissioners or federal appointees who have also served in Congress to develop a measure of agency ideology \citep[see, e.g,][]{Moe_1985,snyder2000shared,nixon2004separation}. However, the votes of commissioners does not reflect how bureaucrats of the agency would have necessarily voted, and the number of federal appointees who have also served in Congress is small.

By far the most common way to measure the ideology of federal agencies have been perception-based measures. Some have analyzed the election contribution data of federal bureaucrats to estimate the ideology of agencies \citep{chen_johnson_2015}. However, individuals must self-identify as an employee of a U.S. federal agency, and there is a self-selection effect with who chooses to donate to candidates during elections. \citet{Clinton_Lewis_2008}, \citet{clinton2012separatedpowers}, and \citet{richardson_clinton_lewis_2018} fielded surveys to both career and appointed bureaucrats across federal agencies. Specifically, \citet{richardson_clinton_lewis_2018} used the 2014 Survey on the Future of Government Service, a survey completed by over 1,500 federal executives. One question in the survey asked respondents whether the policy views of certain agencies were liberal, conservative, or neither. \citet{richardson_clinton_lewis_2018} call these scores the perceived ideological leanings of federal agencies because this measure uses insider knowledge and perceptions of agencies. Aside from measures estimated using LLMs, this is the most up-to-date measure of federal agency ideology.

\subsection{Measuring Latent Attributes Using LLMs}
There is also an emerging literature on using LLMs for estimating measures of latent attributes. Much of this work has focused on estimating the latent attributes of text, such as the level of aversion expressed to Republicans and Democrats in tweets \citep[see, e.g.,][]{cgcot_wu_2024,Le_Mens_Gallego_2025,sarkar-etal-2025-pairscale}. Other works have directly queried information about the politics and positions of specific political actors. These papers leverage the extensive political information embedded in LLMs through their training on massive corpora of digital media, capturing position-taking and widely-held perceptions of political actors. Although the LLM embeds information about the actors' behaviors, these approaches are much closer to perception-based measures than behavior-based measures. \citet{ohagan2024measurementagellmsapplication} prompted GPT to position senators along the ideology dimension, finding that this measure highly correlated with the first dimension of NOMINATE. \citet{wu2023largelanguagemodelsused} used pairwise comparisons between senators to position senators along an ideology dimension, a support for gun control dimension, and a support for abortion rights dimension using GPT and two Llama models of different sizes \citep{grattafiori2024llama3herdmodels}. The outcomes of these pairwise comparisons were scaled using the Bradley-Terry model. \citet{napolio_nd} used the framework from \citet{wu2023largelanguagemodelsused} and made pairwise comparisons between federal executive agencies using GPT, showing that it highly correlates with existing measures of federal agencies. While this current paper builds on \citet{napolio_nd}, the findings of that paper are not reproducible because it used an unknown version of GPT-4.0 that may no longer be available. \citet{napolio_nd} does not propose other estimates of latent attributes of federal executive agencies besides liberal-conservative ideology.

\section{Measuring the Ideology of Federal Agencies Using Various LLMs}

This section describes how we estimated an updated measure of the ideology of federal agencies.  

\subsection{Methodology}
We used the language model pairwise comparison approach described in \citet{wu2023largelanguagemodelsused}. We prompted the three generative LLMs with pairwise comparisons of federal agencies along a liberal-conservative dimension. Pairwise comparisons have many advantages: they only have one task for the LLM to complete, they do not depend on an unknown scale of the LLM when generating numbers, and they can be used with probability models to estimate continuous scores for all items being compared \citep{carlson_montgomery_2017,narimanzadeh2023crowdsourcing}. Pairwise comparisons have also been extensively used in political science \citep{LOEWEN2012212,carlson_montgomery_2017,park2021politiciansgrandstanding} We used the following pairwise comparison prompt across the three LLMs: \textsc{Which agency is perceived to be more liberal: [agency 1] or [agency 2]?}

We then used the Bradley-Terry model \citep{bradleyterry1952} with the outcomes of the pairwise comparisons to estimate ideology scores for each federal agency. The Bradley-Terry model estimates a parameter for each actor being compared. When two actors $i$ and $j$ are compared, the log odds that actor $i$ is ``preferred'' (e.g., more liberal) over actor $j$ is $\text{logit}\left(\Pr(i>j) \right) = \log\left(\frac{\Pr(i>j)}{\Pr(j>i)}\right) = \lambda_i - \lambda_j$. 
Reference-free quasi-standard errors can also be calculated \citep{quasivariances_firth_demenezes}. We called these scores the Agency Ideology Pairwise Scores (henceforth referred to as AIPS). Ties were allowed: a 0.5 ``win'' was added to both agencies when a tie occurred \citep{turner_firth_2012}. 

\subsection{Data}

We used the list of federal agencies from \citet{Bonica2025}, a dataset derived from \citet{richardson_clinton_lewis_2018}.\footnote{The dataset can be found \href{https://docs.google.com/spreadsheets/d/1l80SG27cxUOAHknYvmp-LV6Ag5phGlBVJbqG2U066AI/edit?gid=0\#gid=0}{here}.} The dataset contains 123 total federal agencies and their respective perceived ideology estimates from \citet{richardson_clinton_lewis_2018}. The data also includes an indicator variable that shows which agencies experienced DOGE layoffs.\footnote{Sources for each DOGE layoff can be found in the original dataset.} \citet{Bonica2025} used an indicator variable rather than a count of layoffs per agency because reliable layoff figures were often unavailable. Additionally, layoffs can take various forms, including dismissals or indefinite leaves, making precise numerical estimates difficult to determine. Layoffs were included if they involved forced layoffs; we did not count agencies that lost employees who left from taking the ``buyout.'' The public dataset was last updated on February 28, 2025. We manually updated any additional DOGE layoffs as of May 12, 2025. We made pairwise comparisons between all federal agencies, resulting in 7,503 pairwise comparisons. Each pairwise comparison was repeated three times with each LLM.

We used three state of the art LLMs: Meta's Llama 3.3 70B Instruct (70 billion parameters) \citep{grattafiori2024llama3herdmodels}, Mistral Small 3 (24 billion parameters) \citep{mistral_small_3_2025}, and Microsoft's Phi-4 (14 billion parameters) \citep{abdin2024phi4technicalreport}. All are open-weight, instruction-tuned models developed by different organizations with distinct training methodologies. All models were also released before the start of the DOGE layoffs in February 2025. The selection of LLMs with varying parameter counts and training approaches allows us to examine how estimated agency ideologies might differ across LLMs.

\subsection{Agency Ideology Pairwise Scores}
The Agency Ideology Pairwise Scores for Llama 3.3 70B Instruct are illustrated on the x-axis in Figure~\ref{fig:llama_aips}. From a perspective of face validity, the results made sense: the most liberal federal agencies were the Environmental Protection Agency, Equal Employment Opportunity Commission, National Labor Relations Board, Consumer Financial Protection Bureau, and the Peace Corps. The most conservative federal agencies were the United States Immigration and Custom Enforcement (ICE), Joint Chiefs of Staff, Department of the Army, United States Customs and Border Patrol, and the Missile Defense Agency. The full estimated ideological positions of all agencies for all LLMs with quasi-standard error-derived 95\% confidence intervals can be found in the Appendix.

\subsubsection{Correlations}
We used the Pearson correlation coefficient to compare estimated AIPS across the different LLMs. We also compared each AIPS with the perceived ideology estimate of \citet{richardson_clinton_lewis_2018} (referred to as RCL in the results). The results are in the correlation matrix in Table~\ref{tab:corr_mat}. 

\begin{table}[!ht]
\centering
\caption{Correlations among Agency Ideology Pairwise Scores scores and \citet{richardson_clinton_lewis_2018} (RCL) scores}
\label{tab:corr_mat}
\begin{tabular}{lcccc}
\toprule
                                            & Llama & Mistral & Phi-4  & RCL Score \\ \midrule
Llama                                       & 1.00  & 0.87    & 0.89 & 0.90                                        \\
Mistral                                     &       & 1.00    & 0.92 & 0.80                                        \\
Phi-4                                         &       &         & 1.00 & 0.82                                        \\
RCL Score &       &         &      & 1.00 \\ \bottomrule
\end{tabular}
\end{table}

\begin{figure}[!ht]
    \centering
    \includegraphics[width=\linewidth]{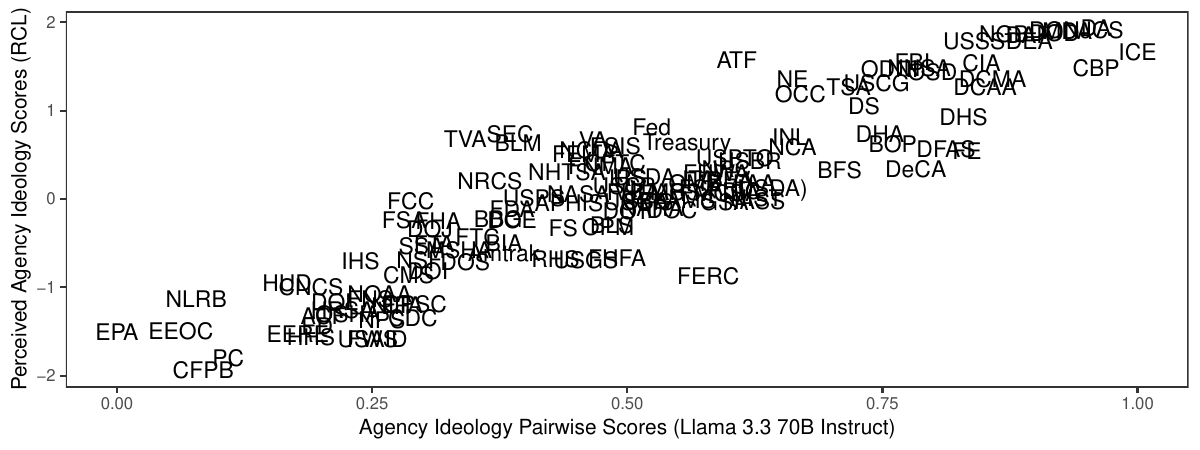}
    \caption{The perceived agency ideology scores from \citet{richardson_clinton_lewis_2018} versus our Agency Ideology Pairwise Scores estimated using Llama 3.3 70B Instruct.}
    \label{fig:llama_aips}
\end{figure}

The correlations between AIPS estimated using different LLMs were high, especially between those estimated using Mistral Small 3 and Phi-4. This indicates substantial agreement in the perceived ideological positioning of federal agencies across the LLMs. Furthermore, each AIPS strongly correlated with the perceived ideology estimate of \citet{richardson_clinton_lewis_2018}. The largest model in this study, Llama 3.3 70B Instruct, had the highest correlation with this measure. Interestingly, the smaller Phi-4 had a slightly higher correlation with the perceived ideology estimate than Mistral Small 3. Figure~\ref{fig:llama_aips} illustrates in greater detail the relationship between AIPS estimated using Llama and \citet{richardson_clinton_lewis_2018}'s perceived agency ideology scores.

\subsubsection{Replicating Bonica [2025] Using the Agency Ideology Pairwise Scores}
We used AIPS to replicate \citet{Bonica2025}'s finding that the perceived ideology of the agency predicts whether the agency experienced DOGE layoffs, controlling for the agency's total budget and number of staff. The results are in Table~\ref{tab:doge}. Using the same logistic regression configuration as \citet{Bonica2025}, we found that the three AIPS replicated the findings in \citet{Bonica2025}, including the finding that annual budget and total staff are not statistically significant predictors of which federal agencies experienced DOGE layoffs.

\begin{table}[!ht] 
\centering 
  \caption{Logistic regression results using Agency Ideology Pairwise Scores estimated using the three LLMs. The outcome is a binary variable for whether the agency experienced DOGE layoffs.} 
  \label{tab:doge} 
\begin{tabular}{@{\extracolsep{2pt}}lccc} 
\toprule
 & Llama & Mistral & Phi-4 \\ 
\midrule \\[-1.8ex] 
Ideology & $-$5.55$^{***}$ & $-$4.57$^{***}$ & $-$5.24$^{***}$ \\ 
         & (1.23) & (1.21) & (1.26) \\ 
log(Annual Budget) & 0.24 & 0.24 & 0.18 \\ 
         & (0.15) & (0.14) & (0.15) \\ 
log(Total Staff) & 0.29 & 0.19 & 0.23 \\ 
         & (0.18) & (0.16) & (0.17) \\ 
Constant & $-$5.74$^{*}$ & $-$4.35 & $-$3.17 \\ 
         & (2.65) & (2.64) & (2.74) \\ 
\midrule
\textit{Note:}  & \multicolumn{3}{r}{$^{*}$p$<$0.05; $^{**}$p$<$0.01; $^{***}$p$<$0.001} \\ 
\end{tabular} 
\end{table}

\section{Measuring the Perception of Federal Agencies as Knowledge Institutions Using an LLM}
Scholars have noted that several figures of the alt-right, such as Curtis Yarvin (also known as ``Mencius Moldbug'') and Bronze Age Pervert (``BAP''), appear to be influential on current actions of the American federal government. Vice President JD Vance, for example, has directly referenced Yarvin's ideas in speeches and podcasts \citep{Wilson2024Yarvin}. These figures of the alt-right, commonly coined neo-reactionaries, have been described as part of a broader anti-democratic, anti-egalitarian, and reactionary movement \citep{Jones2019}.

One of Yarvin's core ideas is the ``Cathedral,'' which \citet{Jones2019} describes as the ``superstructure of cultural capital within universities, the media, and bureaucracy, which it views as not only hegemonic and inefficient, but also the primary reason for the decline of Western civilization, as it has embraced liberal humanism.'' In other words, the Cathedral is problematic for Yarvin because these knowledge institutions, which create, legitimize, and distribute knowledge, have been taken over by progressive messaging and need to be dissolved \citep{Marchese2025Yarvin}.

We estimated a ``knowledge institution'' measure using the same pairwise comparison approach and data described in the previous section. We used the following prompt: \textsc{Knowledge institutions create, distribute, and/or legitimize knowledge. Which agency is more likely to be perceived to be producing knowledge, distributing knowledge, and/or supporting knowledge institutions such as academic and educational institutions, the media, and civil society organizations: [Agency A] or [Agency B]?}

We used Llama 3.3 70B Instruct, the largest LLM among the three used in the previous section and the LLM with the highest correlation with the \citet{richardson_clinton_lewis_2018} perceived ideology measure. We called this measure the Knowledge Institution Pairwise Scores (henceforth referred to as KIPS). No similar measure has been previously estimated in this literature.

\begin{figure}[!ht]
    \centering
    \includegraphics[width=\linewidth]{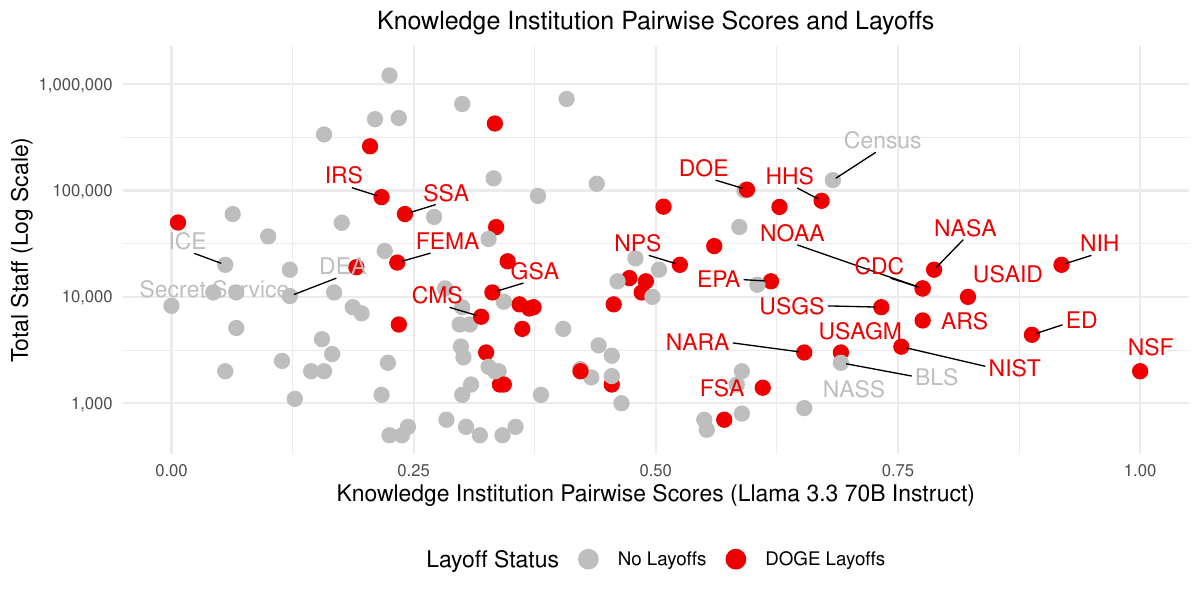}
    \caption{Total staff by agency (on a log scale) versus our Knowledge Institution Pairwise Scores estimated using Llama 3.3 70B Instruct.}
    \label{fig:kp_plot}
\end{figure}

The resulting measure can be seen in Figure~\ref{fig:kp_plot}. The figure shows that KIPS did not correlate with total staffing at the agency; yet, agencies with higher KIPS appeared to be disproportionately hit with layoffs. The agencies with the highest KIPS are the National Science Foundation, the National Institutes of Health, the Department of Education, the United States Agency for International Development (supporting education and civil societies worldwide and funding research on international development), and NASA. The full estimated positions of agencies' perceptions as knowledge institutions with quasi-standard error-derived 95\% confidence intervals can be found in the Appendix. 

We confirmed this analysis using logistic regressions in Table~\ref{tab:logistic_kips}. We used three different model configurations. Model 1 included only KIPS; Model 2 included KIPS and AIPS using Llama; Model 3 included KIPS and the perceived agency ideology scores from \citet{richardson_clinton_lewis_2018}. All models included controls for annual budgeting and total staffing.

\begin{table}[!ht] 
  \centering 
  \caption{Logistic regression results using three different model configurations. The outcome was a binary variable for whether the agency experienced DOGE layoffs.} 
  \label{tab:logistic_kips} 
\begin{tabular}{@{\extracolsep{2pt}}lccc} 
\toprule
 & Model 1 & Model 2 & Model 3 \\ 
\midrule \\[-1.8ex] 
Knowledge Institution Pairwise Scores & 5.00$^{***}$    & 3.21$^{*}$     & 2.98$^{*}$ \\
                                                        & (1.21)          & (1.34)         & (1.34) \\
Agency Ideology Pairwise Scores                         &                 & $-$4.27$^{**}$ & \\
                                                        &                 & (1.33)         & \\
Perceived Agency Ideology \citep{richardson_clinton_lewis_2018}                                            &                 &                & $-$1.20$^{***}$\\ 
                                                        &                 &                & (0.34)\\
log(Annual Budget)                                      & 0.31$^{*}$      & 0.25           & 0.24 \\ 
                          & (0.14)          & (0.15)         & (0.15) \\ 
log(Total Staff)          & 0.02            & 0.25           & 0.33 \\ 
                          & (0.15)          & (0.18)         & (0.19) \\ 
Constant                  & $-$9.71$^{***}$ & $-$7.65$^{**}$        & $-$10.14$^{***}$ \\ 
                          & (2.67)          & (2.83)         & (2.87) \\ 
\midrule
\textit{Note:}  & \multicolumn{3}{r}{$^{*}$p$<$0.05; $^{**}$p$<$0.01; $^{***}$p$<$0.001} 
\end{tabular} 
\end{table}

KIPS were a significant predictor of DOGE layoffs, even when holding agency ideology---whether it was estimated using an LLM or previous expert surveys---constant. This empirically suggests that DOGE layoffs were driven by more than just ideology: they were also driven by institutional factors related to the perception of knowledge production and management within agencies.

\section{Using LLMs in Place of Expert Political Surveys: Advantages, Alternative Views, and a Two-Part Theory-Gap Criterion}

\subsection{Advantages}
The DOGE layoffs case study demonstrated that there are several advantages to using LLMs in place of expert political surveys. First, and most importantly, this measurement approach offers \textbf{temporal flexibility}: it allow us to measure latent attributes of political actors using approaches that would be difficult or impossible with experts. Surveys capture the now, while LLMs allow us to capture an ex ante snapshot that no expert panel can retrospectively supply. Using open weight models with verified checkpoint dates preceding the shock guarantees that the model's knowledge cutoff is pre-event, preventing post-event biases from contaminating the analysis.

This temporal flexibility also demonstrates one fundamental difference with the literature on using language models to serve as surrogate, virtual subjects, sometimes referred to as ``silicon samples'' \citep[see, e.g.,][]{Argyle_Busby_Fulda_Gubler_Rytting_Wingate_2023,dillion2023canailm,Bisbee_Clinton_Dorff_Kenkel_Larson_2024,moon-etal-2024-virtual,tjuatja2024llmshuman,sarstedt2024siliconsamples,suh2025languagemodelfinetuningscaled}. Many of these works analyze whether LLMs are able to replicate human responses on surveys and if these patterns generalize to unseen surveys. These approaches are inherently forward-looking: they focus on using information from the past that the LLM was trained on to predict future responses, behaviors, and affect. Our argued position on using LLMs for estimating latent attributes is, by contrast, retrospective: we leverage pre-shock information and perceptions embedded in LLMs to construct measures that help us better understand the associated factors in observed outcomes.

LLMs also allow us to estimate measures in a \textbf{speedy and cost efficient manner}. This speed is often critical to understanding rapidly changing circumstances. Moreover, when traditional measurement approaches are still, or become, available, using LLMs can act as a way to pilot a study, flagging ambiguous items, estimating effect sizes and variances, and guiding sample size and questionnaire design so that subsequent expert or public surveys are faster, cheaper, and more focused. Relatedly, the \textbf{scalability} of LLMs enables researchers to test all combinations of items using measurement approaches such as pairwise comparisons. It allows us to construct a preliminary measure when surveys are not feasible, when the shock is still unfolding, or when human fatigue from completing lengthy surveys increases in the aftermath of the shock \citep{porter2004multiplesurveys,Jeong2023ExhaustiveFatigue}. Lastly, LLMs have a \textbf{more uniform knowledge base}: experts are often familiar with a subset, but not all, federal agencies. For example, in the survey that \citet{richardson_clinton_lewis_2018} fielded, the respondents were asked about the three agencies they worked with the most along with a set of randomly chosen agencies. LLMs can potentially be used to fill knowledge gaps where experts lack familiarity with certain agencies.

\subsection{Alternative Views}
The most important alternative view to consider would be to \textit{not} use LLMs to estimate latent attributes even when a population can no longer be surveyed. This view could be justified by several potential disadvantages with this approach. We focus on three potential disadvantages.

The model's responses to pairwise comparisons may be \textbf{biased}---they are inherently constrained by the LLM's training data. If an agency is underrepresented in the training data, the model may systematically misestimate their latent attributes. Training data is often unavailable for auditing, even for open weight models such as Llama \citep{grattafiori2024llama3herdmodels}. Moreover, training data do not capture the insider knowledge that many experts utilize when completing surveys.

Relatedly, it is \textbf{difficult to quantify the uncertainty} of a model's generated response \citep[see, e.g.,][]{baan2023uncertaintynaturallanguagegeneration,lin2024generating,liu2024uncertaintyestimationquantificationllms}. This may be particularly relevant when, for example, two relatively obscure agencies are compared. While experts can describe their uncertainty in responses, LLMs cannot do so directly, although there are approaches that could potentially calibrate confidence metrics \citep{mielke-etal-2022-reducing,lin2022teaching,zhou-etal-2023-navigating,xiong2024can,Farquhar2024}. Moreover, there are the usual concerns about model sensitivity to hyperparameters and prompt construction \citep[see, e.g.,][]{sclar2024quantifying,zhuo-etal-2024-prosa,wu2025semanticuncertainty}. 

Perhaps most importantly, there is often \textbf{limited external validation for new constructs} \citep{Adcock_Collier_2001}. For example, we do not have a benchmark against which to compare KIPS. At the same time, we cannot use expert surveys now because of post-event bias. While there may be indirect validity checks, such as examining agency behavior around collaborating with universities and civil societies, they are not often correlated with the \textit{perception} of specific agencies as knowledge institutions. An approach could be to survey experts now, despite the caveats described above, providing an imperfect form of validation. What this imperfect validation would tell us about the LLM-driven measures is beyond the scope of this paper.

Our position is that LLMs can be used to recover aggregate perceptions about agencies with respect to theoretically backed latent attributes. Our aim is \textit{not} to anthropomorphize models or simulate individual experts completing political surveys. Iterative prompt refinements could yield spurious latent constructs that appear to correlate with real-world actions despite having no meaningful theoretical basis.

\subsection{A Two-Part Theory-Gap Criterion}
To mitigate these concerns beyond using the typical best practices for using LLMs in computational social science settings \citep{ziems2024llmscss}, we propose a two-part theory-gap criterion for when researchers could use an LLM in place of expert political surveys. First, there must be a \textbf{substantive theory}: there must be strong theory or real-world evidence that the construct should matter for the outcome of interest. Second, there must be a \textbf{measurement gap}: there must be a reason why expert survey data cannot be collected, and it must have been theoretically possible to collect this data before the shock. Survey data collection barriers may stem from temporal constraints, such as a rapidly unfolding event that cannot wait for expert surveys to be collected, or from post-event knowledge that could bias downstream inferences about the event. If the population was inaccessible pre-shock, the LLM likely lacks sufficient training data to provide reliable information about the measure of interest.

The DOGE layoffs case study meets both criteria. There are substantive theories justifying why both ideology and perception as knowledge institutions matter, such as Trump saying certain agencies were ``run by a bunch of radical lunatics'' \citep{hirsh2025elon} and Vance citing Yarvin's theories. Although universities are not agencies, withdrawing federal funding from colleges and halting Harvard from enrolling international students signal the Trump administration's stance toward knowledge institutions more broadly \citep{nyt2025harvard}. It also satisfies the measurement gap criterion: pre-event perceptions necessary for analyzing the associative factors cannot be obtained through expert surveys. Moreover, since bureaucrats have successfully participated in surveys previously, it would have been feasible to administer a survey assessing perceptions of agencies as knowledge institutions.

\section{Conclusion}
We take the position that LLMs can be a viable substitute for expert political surveys when a shock disrupts traditional measurement approaches. Using the 2025 DOGE layoffs as a case study, we show that we can provide empirical evidence that certain theories and observed patterns hold true, such as DOGE targeting federal agencies both along ideological and perceptions as knowledge institutions dimensions. We propose a two-part theory-gap criterion that we suggest practitioners follow when using LLMs in place of expert political surveys. Although we focused on DOGE layoffs in this position paper, this approach can be generalized to other shocks such as pandemics (e.g., public trust in health agencies), market crashes (e.g., firm-risk perceptions), and geopolitical events (e.g., credibility of state actors). 

\newpage

\bibliographystyle{plainnat}
\bibliography{bibliography}

\newpage
\appendix
\section{Appendix}

\subsection{Agency Ideology Pairwise Scores by LLM}

\begin{figure}[!ht]
    \centering
    \includegraphics[width=\linewidth]{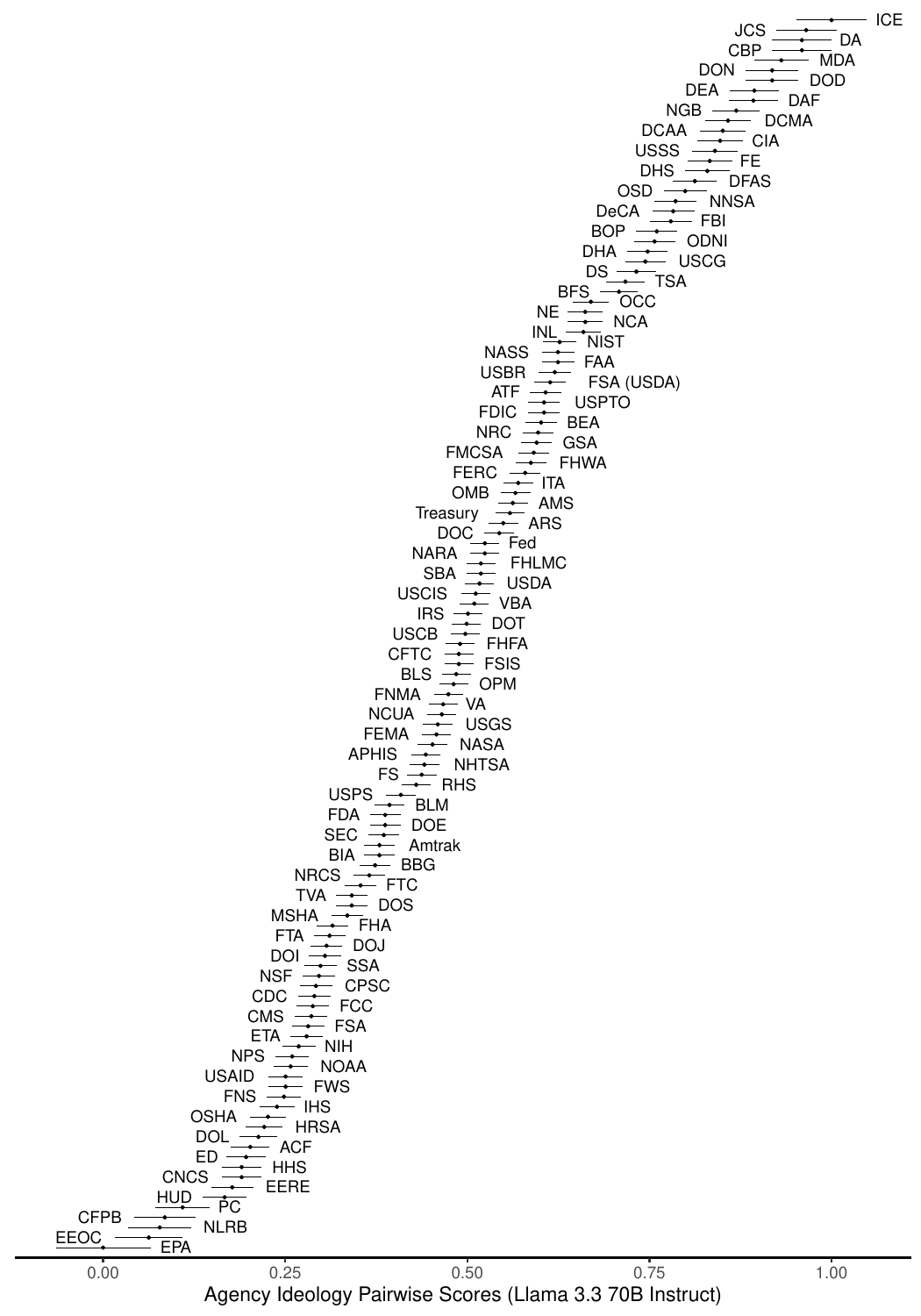}
    \caption{AIPS for Llama 3.3 70B Instruct. 95\% confidence intervals are based on quasi-standard errors.}
    \label{fig:AIPS_Llama}
\end{figure}

\begin{figure}[!ht]
    \centering
    \includegraphics[width=\linewidth]{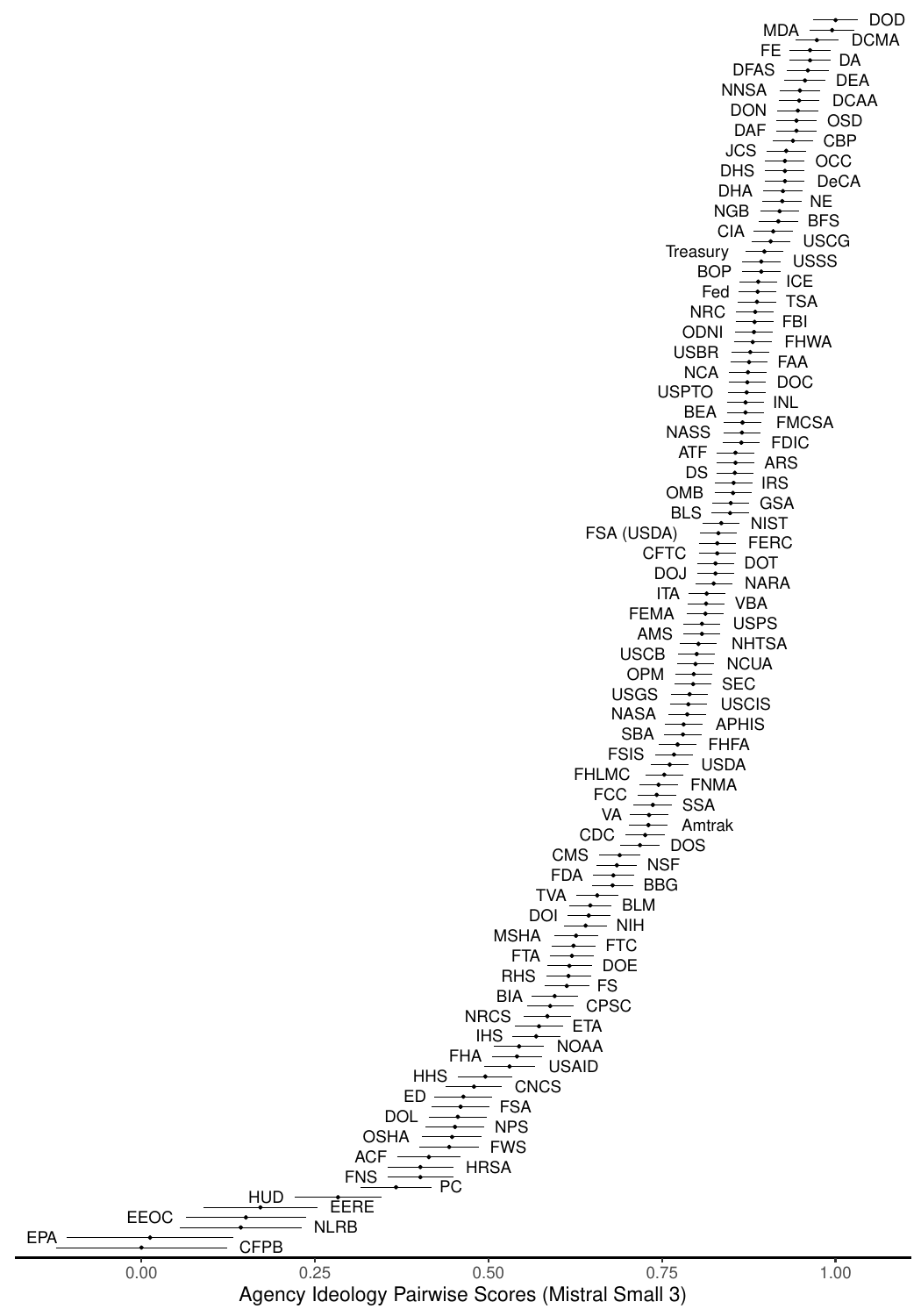}
    \caption{AIPS for Mistral Small 3. 95\% confidence intervals are based on quasi-standard errors.}
    \label{fig:AIPS_Mistral}
\end{figure}

\begin{figure}[ht!]
    \centering
    \includegraphics[width=\linewidth]{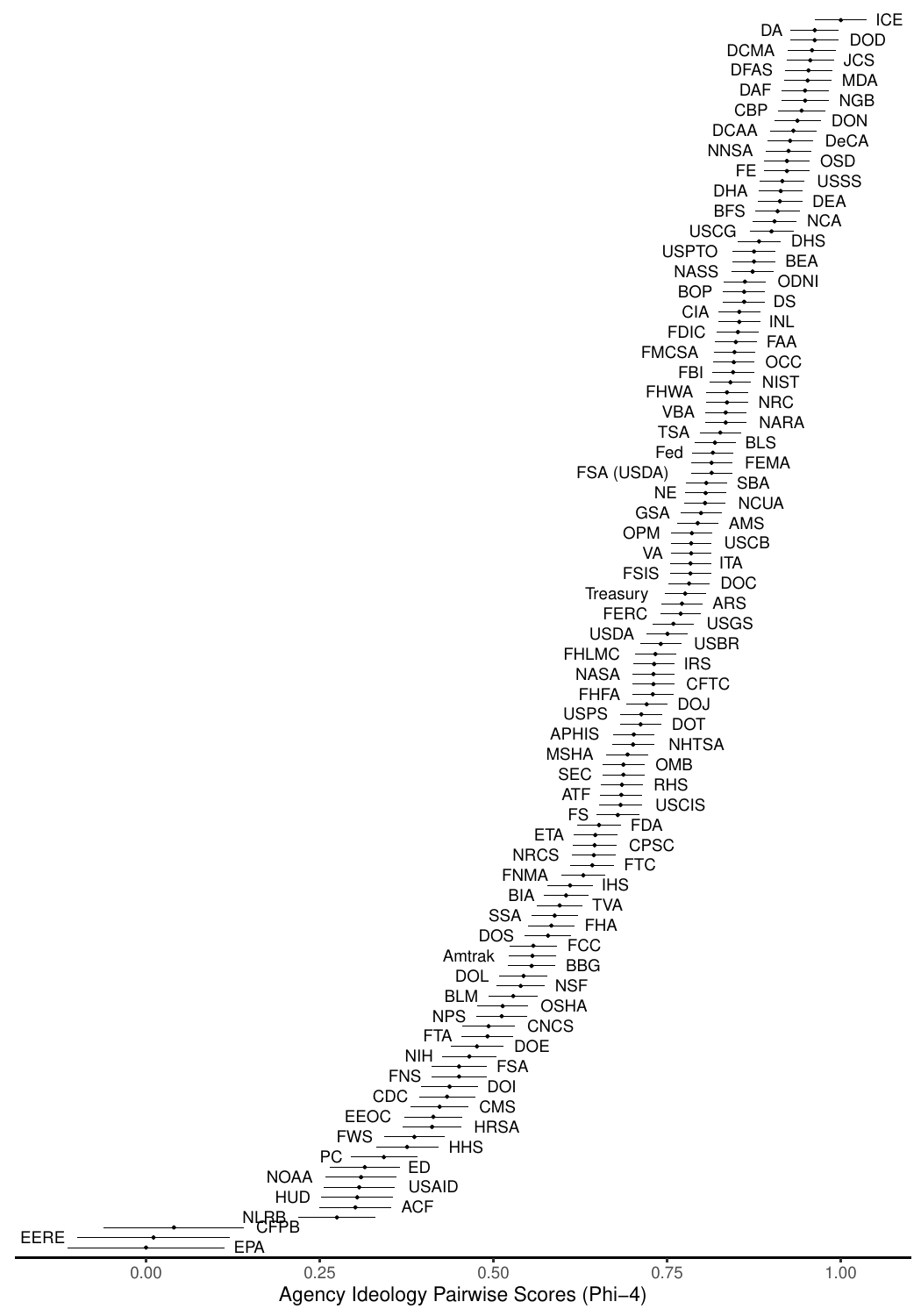}
    \caption{AIPS for Phi-4. 95\% confidence intervals are based on quasi-standard errors.}
    \label{fig:AIPS_Phi}
\end{figure}

Figure~\ref{fig:AIPS_Llama} shows the AIPS for Llama 3.3 70B Instruct. Figure~\ref{fig:AIPS_Mistral} shows the AIPS for Mistral 3 Small. Lastly, Figure~\ref{fig:AIPS_Phi} shows the AIPS for Phi-4.

\subsection{Knowledge Institution Pairwise Scores}

\begin{figure}[!ht]
    \centering
    \includegraphics[width=\linewidth]{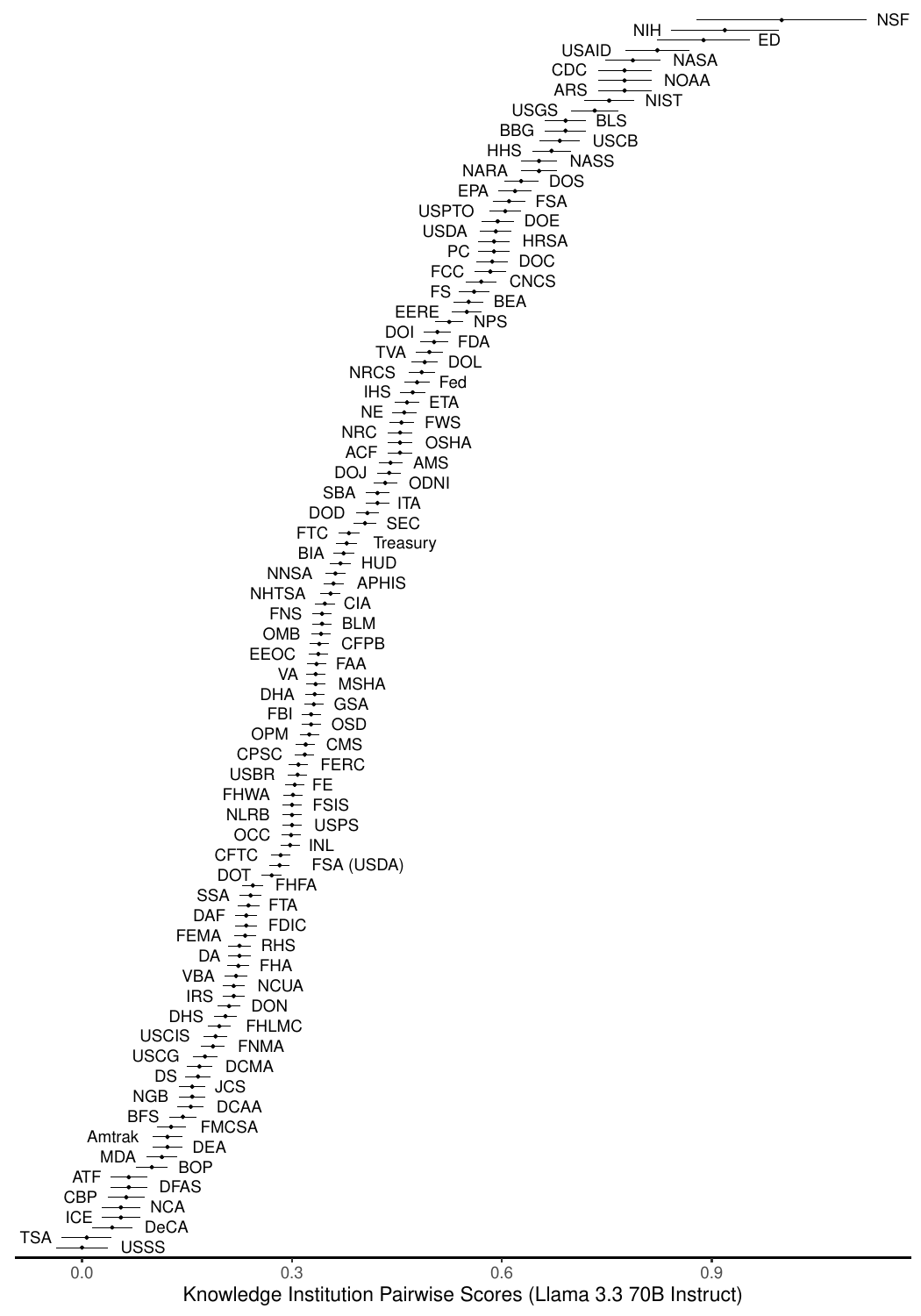}
    \caption{KIPS for Llama 3.3 70B Instruct. 95\% confidence intervals are based on quasi-standard errors.}
    \label{fig:KIPS_Llama}
\end{figure}

Figure~\ref{fig:KIPS_Llama} shows the KIPS for Llama 3.3 70B Instruct.

\subsection{Extracting Answers from Models' Responses}
\citet{rottger-etal-2024-political} demonstrated that LLMs can give different answers when forced to respond in a specific format. We did not constrain or force the LLM to respond in a specific format. Instead, we employed a follow-up prompt to extract the answer from the model's unconstrained response.

The follow-up extraction prompt for Agency Ideology Pairwise Scores was the following:
\begin{quote}
    \textsc{Using just your answer above, which agency is perceived to be more liberal: the `[agency 1]' or the `[agency 2]'? Only return the full name of the agency, with no other words. If neither agencies are more liberal or more conservative, return ``Tie'' with no other words or punctuation.}
\end{quote}

The follow up extraction prompt for Knowledge Institution Pairwise Scores was the following:
\begin{quote}
    \textsc{Using just your answer above, which agency is more likely to be perceived to be producing knowledge, distributing knowledge, and/or supporting knowledge institutions such as academic and educational institutions, the media, and civil society organizations: the `[agency 1]' or the `[agency 2]'? Only return the full name of the agency, with no other words. If neither agencies are more or less likely to be perceived to be producing knowledge, distributing knowledge, and/or supporting knowledge institutions, return ``Tie'' with no other words or punctuation.}
\end{quote}

\subsection{LLM Hyperparameters}
When making pairwise comparisons, we used the default temperature and top-$p$ hyperparameters for all LLMs. During the answer extraction phase, we set the temperature to 0.

\end{document}